\documentclass[aps,prd,preprint]{revtex4}
\usepackage{amsfonts, amssymb, amsmath, stmaryrd, 
cancel, mathrsfs, mathcomp, calrsfs}
\usepackage{graphicx} 

\usepackage[english]{babel}
\usepackage[latin1]{inputenc}

\begin{document}
\title{Many-quark interactions: Large-$N$ scaling and contribution to baryon masses}

\author{Fabien~Buisseret}%
 \email{e-mail: fabien.buisseret@umons.ac.be, ORCiD: 0000-0002-7801-036X}
\affiliation{CeREF Technique, 159 Chauss\'{e}e de Binche, 7000 Mons, Belgium}
\affiliation{
Service de Physique Nucl\'{e}aire et Subnucl\'{e}aire, Universit\'{e} de Mons, UMONS Research Institute for Complex Systems, 20 Place du Parc, 7000 Mons, Belgium}

\author{Cintia~T.~Willemyns}%
 \email{e-mail: cintia.willemyns@umons.ac.be, ORCiD: 0000-0001-8114-0061}
\affiliation{Service de Physique Nucl\'{e}aire et Subnucl\'{e}aire, Universit\'{e} de Mons, UMONS Research Institute for Complex Systems, 20 Place du Parc, 7000 Mons, Belgium}

\author{Claude~Semay}%
 \email{e-mail: claude.semay@umons.ac.be, ORCiD: 0000-0001-6841-9850}
\affiliation{Service de Physique Nucl\'{e}aire et Subnucl\'{e}aire, Universit\'{e} de Mons, UMONS Research Institute for Complex Systems, 20 Place du Parc, 7000 Mons, Belgium}


\begin{abstract}
Starting from an effective Hamiltonian modelling a baryon made of $N$ identical quarks in the large-$N$ approach of QCD, we obtain analytical formulas allowing to estimate the contributions of multiquark interactions to the baryon mass. The cases of vanishing (mass spectrum) and non-vanishing (baryon melting) temperatures are treated.  

\end{abstract}

\maketitle

\section{Introduction}\label{sec:Intro}

If quantum chromodynamics (QCD) is indisputably the theory of the strong interaction, many different approaches must be explored to study this theory in the domain of the hadron spectroscopy. Among the most efficient and popular ones are the constituent models and the large-$N$ picture. In the first models, constituent quarks appear like ``dressed" current quarks and interact via potentials simulating the exchanges of virtual gluons and quark-antiquarks pairs \cite{isgur}. In the large-$N$ picture, the SU(3) group of QCD is replaced by SU($N$), allowing an expansion for the properties of the theory in powers of $1/N$, which is treated as a small parameter \cite{witten}. It appears quite natural to try to combine these two different methods to gain new insights about the hadron physics. Several works of this type have de facto obtained some interesting results, but with only one- and two-body interactions \cite{Buisseret:2010na,Buisseret:2011aa,Willemyns:2016}. In this work, such a methodology is used to study multiquark interactions within the baryon. Note that various large-$N$ versions of QCD exist in which quarks can be in different colour representations \cite{Buisseret:2010na,Buisseret:2011aa}; we here consider quarks in the fundamental representation, \textit{i.e.} 't Hooft limit \cite{tHooft:1973alw}. 

The ground state mass of a baryon made of $N$ quarks of the same flavour can a priori be expressed as 
\begin{equation}\label{M0}
M=N K+\sum^{h_\textrm{max}}_{h=1} \begin{pmatrix} N \\ h \end{pmatrix} E_h , 
\end{equation} 
where $K$ is the average kinetic energy of one quark (including its rest energy) and where $E_h$ is the average potential energy of an interaction involving $h$ quarks, this energy being weighted by the binomial coefficient $\begin{pmatrix} N \\ h \end{pmatrix}$. 

It has been said that three (or more)-quark interactions ``simply renormalise the average Hartree potential" in a mean field approximation for two-body interactions only \cite{witten}: The necessary technical complications for their inclusion in a quark model can thus be avoided, in principle, since they can be absorbed in a redefinition of the parameters. Nevertheless, it has been shown in \cite{Dmitrasinovic:2001nu} for $N=3$ that three-quark interactions going beyond OGE have relevant physical features such as increasing the mass gap between colour singlet baryonic states and unphysical coloured ones, or lifting mass degeneracies in tetraquarks. Only a few works have addressed the problem of three-quarks interactions in quark models \cite{Dmitrasinovic:2001nu,Pepin:2001is,Papp:2002xh}, while, to our knowledge, the question of multiquark interactions in large-$N$ QCD has only been studied in \cite{Albertus:2015fca}. In this last work, it is concluded that $h$-quark interactions ($h>2$) are suppressed by a relative factor $h!$ and can then be neglected. This result is obtained with a wave function computed in the Hartree approximation, as in \cite{witten}. We want to reconsider this problem by using the envelope theory (ET), which is a method to compute approximate but reliable solutions for many-body quantum systems \cite{Semay:2013}. It is quite easy to implement if all the particles are identical and the computation cost is independent of the number of particles. Moreover, analytical upper or lower bounds can be obtained in favourable situations \cite{Semay:2015}. This will be the case for the systems considered here.

The Hamiltonian for baryons in the large-$N$ approach is given in Sec.~\ref{sec:barmass}. The dependence on $N$ for the parameters are fixed in Sec.~\ref{sec:lnc}. One-gluon exchange and three-quark interactions are respectively treated in Sec.~\ref{sec:oge} and Sec.~\ref{sec:3qi}. While in these sections, baryons are always considered at vanishing temperature, their behaviour above the deconfinement temperature $T_c$ is examined in Sec.~\ref{sec:deconf}. Concluding remarks are given in Sec.~\ref{sec:conclu}.

\section{Baryon mass formula}\label{sec:barmass}


We aim at deriving the mass formula (\ref{M0}), which is based on intuitive mean-field arguments, from an effective theory based on constituent quarks. A generic quark model of baryon made of $N$ identical quarks at vanishing temperature, including one- and $h$-quark interactions may be defined by the Hamiltonian
\begin{equation}\label{Hamdef}
	H=\sum^N_{i=1}\Big[ K(\vert \vec p_i\vert)+U(\vert\vec x_i-\vec R\vert)\Big]	+\sum^N_{\{i_1,\dots,i_h\}} V_h(r_{\{i_1,\dots,i_h\}})
\end{equation}
with $K$ the kinetic energy, $U$ the one-body interaction simulating the confinement (with $\vec R$ being the centre of mass position) and $V_h$ a $h$-body potential. $\vec x_i$ and $\vec p_i$ are the individual conjugate variables. The $h$-body variables are defined by $r_{\{i_1,\dots,i_h\}}=\sqrt{\sum^{\{i_1,\dots,i_h\}}_{i<j} r^2_{ij}}$ and $r_{ij}=\vert \vec x_i-\vec x_j\vert$. The symbol $\{i_1,\dots,i_h\}$ denotes a set of $h$ quarks among the $N$ quarks in the baryon, with $i_1<\dots<i_h$. The sum $\sum^N_{\{i_1,\dots,i_h\}}$ runs over the different sets $\{i_1,\dots,i_h\}$ while the sum $\sum^{\{i_1,\dots,i_h\}}_{i<j}$ runs over the different pairs in a particular set $\{i_1,\dots,i_h\}$. We assume that $h$ is finite, that is not of order $N$. This kind of many-body interaction has already been used in hadronic physics \cite{Dmitrasinovic:2001nu,Pepin:2001is} or for systems of cold atoms \cite{Kievsky:2020}.

As shown in \cite{Semay:2018rvp} by resorting to the ET, an approximate mass formula may be found by solving the following set of equations ($\hbar = c = 1$):
\begin{eqnarray}\label{model0}
	M&=&N\, K\left(\frac{Q}{Nx_0}\right)+N\, U(x_0)+\begin{pmatrix} N \\ h \end{pmatrix} V_h\left(  \psi_h x_0\right), \nonumber \\
\frac{Q}{Nx_0}\, K'\left(\frac{Q}{Nx_0}\right)&=&x_0\, U'(x_0)+\begin{pmatrix} N \\ h \end{pmatrix} \frac{\psi_h}{N} x_0 V'_h\left(  \psi_h x_0\right),
\end{eqnarray}
where a prime denotes a derivative, and with the global quantum number  
\begin{equation}
	Q=\sum^{N-1}_{j=1}\left( 2n_j+\ell_j+\frac{3}{2}\right)
\end{equation}
and where
\begin{equation}
	\psi_h=N\sqrt{\frac{h(h-1)}{N(N-1)}}.
\end{equation}
This formula is of the form (\ref{M0}) as expected, and it is valid for the ground state as well as for the excited states, a particular state being defined by the global quantum number $Q$. The quantum numbers $\{ n_j,\ell_j \}$ are associated with the $N-1$ internal Jacobi coordinates and the form of $Q$ originates from the exact solutions of translation-invariant $N$-body harmonic oscillator Hamiltonians (see \cite{Willemyns:2021} and references therein) on which the ET relies. Quark spin is obviously neglected but, since the singlet colour wave function is totally antisymmetric, quarks can be seen as ``quasi-bosons" with a totally symmetric spatial wave function if the spin-flavour part is also totally symmetric. 

Confinement in baryons has been extensively studied within quark models since the celebrated work \cite{isgur}. They are typically associated with a confinement potential of the form $\sigma \sum^N_{i=1}\vert \vec x_i-\vec x_0\vert$, where $\sigma$ is the string tension and where $\vec x_0$ is the position of the string junction, which can be identified with the center of mass $\vec R$ in good approximation \cite{witten,Silvestre:2003}. Therefore, this justifies the one-body form of the confinement in Hamiltonian~(\ref{Hamdef}) with $U(x)=\sigma\, x$. Explicit $(h\geq 2)$-quark interactions will be discussed later. Our ansatz for $V_h$ is a scaling of the form
\begin{equation}
	V_h= {\cal C}_h\, g_h v_h,
\end{equation}
with $g_h$ a coupling constant a priori depending on the strong coupling constant $g$ (see next section), ${\cal C}_h$ a colour factor involving SU($N$) Casimir operators and $v_h$ the space part of the average potential energy.

\section{Large-$N$ scaling}\label{sec:lnc}  

The large-$N$ behaviour of our model can now be investigated. In the following, the symbol $\sim$ is used for a quantity evaluated for $N\to \infty$ (possibly up to a multiplicative constant). For states with finite quantum numbers $\{ n_j,\ell_j \}$, $Q/N \sim 3/2$ ($Q=3(N-1)/2$ for the fully symmetrical ground state). Moreover, $\psi_h\sim \sqrt{h(h-1)}$. If $h$ is finite, it is useful to recall that $\left( \displaystyle\frac{N}{h}\right)^h\leq \begin{pmatrix} N \\ h \end{pmatrix}\leq \displaystyle\frac{N^h}{h!}$. Hence, $\begin{pmatrix} N \\ h \end{pmatrix}\sim a_h\, N^h$ where $a_h$ does not depend on $N$. 

Assuming the Casimir scaling \cite{Bali:2000,Lucini2001,Semay:2004,Buisseret:2020}, which follows from a strong coupling expansion in lattice QCD, one has $\sigma = \frac{N^2-1}{2N} g^2 \Omega$ with $\Omega$ a constant \cite{Kogut:1979vg,Kogut:1980pm}. Since $g^2=\lambda/ N$ with $\lambda$ the 't Hooft coupling, $\sigma$ is constant at large-$N$. 

The power in $N$ of the coupling constant $g_h$ may be found as follows. Any Feynman diagram involving $h>2$ quarks must contain $h$ quark-antiquark-gluon vertices so that each quark emits a gluon line: a $g^h$ factor is present. Then the quarks have to be ``linked". Linking two quarks by the exchange of one gluon costs at least one quark-antiquark-gluon vertex or a three-gluon vertex with one outgoing gluon line, \textit{i.e.} a factor $g$. There are at least $h-2$ such links, so the dominant Feynman diagrams are proportional to $g^{2h-2}$. One can therefore state that the $h$-quark coupling constant is such that 
\begin{equation}\label{gh}
	g_h\sim g^{2h-2}\equiv \bar g_h\, N^{1-h}
\end{equation}
for the leading-order diagrams. By definition, $\bar g_h$ does not depend on $N$. Notice that two consecutive three-gluon vertices may be replaced by a single four-gluon vertex, which does not change the order in $g$. Equation (\ref{gh}) is thus valid for any leading-order diagram.

Gathering all these results, when $N\to \infty$, equations (\ref{model0}) become
\begin{eqnarray}\label{model1}
	\frac{M}{N}&\sim&K\left(\frac{3}{2x_0}\right)+\sigma x_0+a_h\,{\cal C}_h\, \bar g_h v_h\left(  \sqrt{h(h-1)} x_0\right), \nonumber \\
	\frac{3}{2x_0}\, K'\left(\frac{3}{2x_0}\right)&\sim&\sigma x_0+a_h\, \sqrt{h(h-1)}\, x_0 {\cal C}_h\, \bar g_h v'_h\left(  \sqrt{h(h-1)} x_0\right).
\end{eqnarray}
It follows form equations (\ref{model1}) that ${\cal C}_h$ has to be constant at large-$N$ at leading order for the baryon mass to be of order $N$. We will now check that known explicit interactions are consistent with that requirement. Recall that $\frac{1}{h^h}\leq a_h\leq \frac{1}{h!}$. The fact that $h$-quark one-gluon exchange interactions bring a contribution of order $N/h!$ to the baryon mass was already suggested in \cite{Albertus:2015fca} for heavy baryons. Here we extend this result to arbitrary baryons made of identical quarks and show that the actual suppression is even stronger than $1/h!$.

 \section{One-gluon exchange}\label{sec:oge}   

At leading order in $N$, two-quark interactions are one-gluon exchange (OGE) processes, typically associated with a Coulomb potential of the form 
\begin{equation}
{\cal C}_{2\, \textrm{OGE}} \sum^N_{i<j}\frac{\alpha_s}{\vert \vec x_i-\vec x_j\vert} \quad \textrm{with} \quad {\cal C}_{2\, \textrm{OGE}}=\frac{1}{2}\left(C^2_2-2C^1_2\right),
\end{equation}
where $\alpha_s=\frac{g^2}{4\pi}$ and where $C^k_p$ is the eigenvalue of the SU($N$) Casimir operator of order $p$ for the totally (anti)symmetric representation with $k$ indices. Explicit formulas for $C^k_p$ are given in \ref{sec:app}. 

\begin{figure}
	\includegraphics[width=6cm,height=2.75cm]{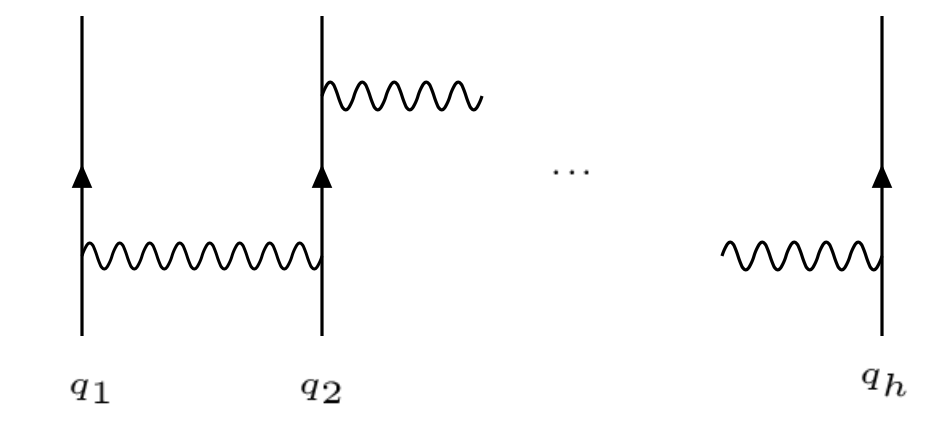}
	\caption{One-gluon exchange process involving $h$ quarks.}
	\label{fig:1}       
\end{figure}

An obvious generalisation to $h$-quark interaction is the multiple OGE sketched in Fig. \ref{fig:1}. For colour singlet baryons, two quarks are always in the antisymmetrical representation and, according to~(\ref{C2OGE}), 
\begin{equation}
	{\cal C}_{h\, \textrm{OGE}}=\left[{\cal C}_{2\, \textrm{OGE}}\right]^{h-1}= \left[-\frac{N+1}{2N}\right]^{h-1}.
\end{equation}
It is of order 1 at large-$N$ as required. 

One- and two-quark interactions bring a contribution of order $N$ to the baryon mass as it has been checked within a Hamiltonian approach with relativistic quark kinetic energy and linear plus Coulomb potential in \cite{Buisseret:2010na}. This results is here extended to OGE exchanges involving $h$ quarks. It is worth mentioning that one-gluon exchange processes also lead to hyperfine corrections (spin-orbit, spin-spin, etc.). It has been shown in \cite{Buisseret:2011aa} that spin-spin corrections only bring a contribution of order $1/N$ to the baryon mass and can then be neglected in a first approximation.

\section{Three-quark interactions}\label{sec:3qi} 

\begin{figure}
	\includegraphics[width=6cm,height=4.30cm]{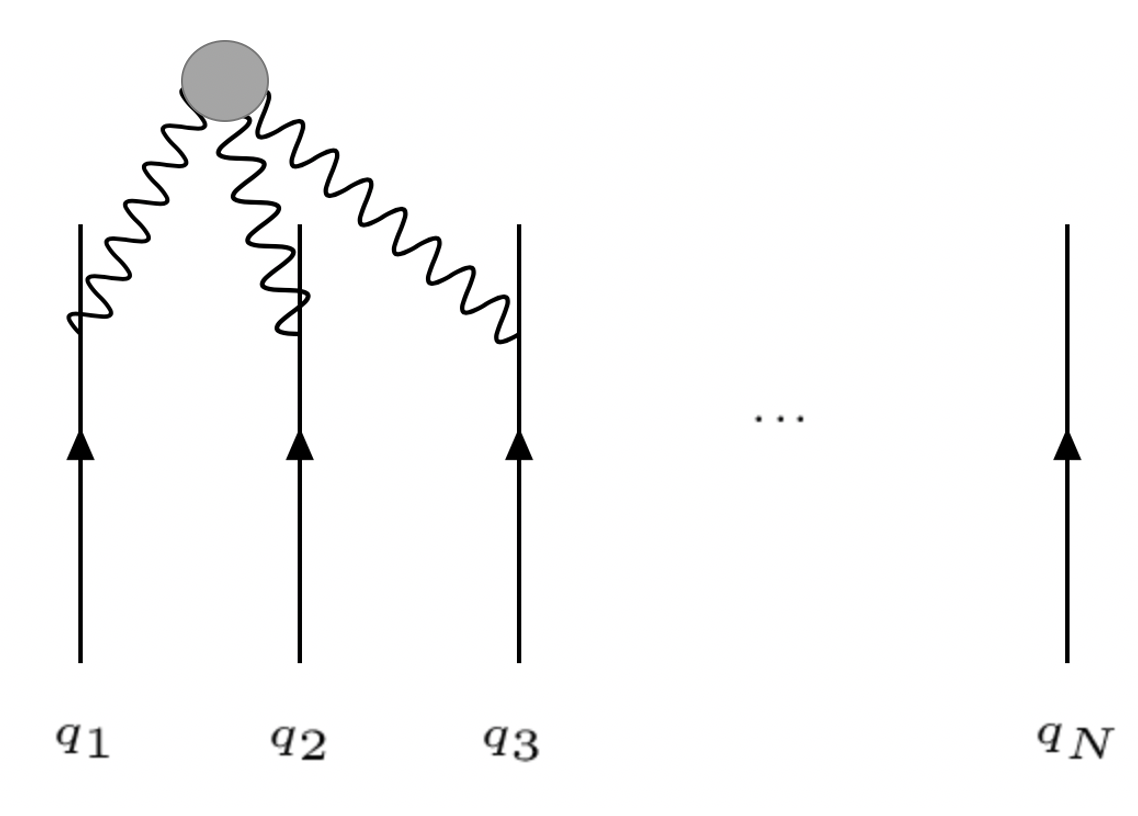}
	\caption{Typical three-quark interaction beyond OGE.}
	\label{fig:2}       
\end{figure}

We now focus on three-quark interactions beyond OGE; an example is shown in  Fig. \ref{fig:2}. The natural choice made in \cite{Dmitrasinovic:2001nu,Pepin:2001is} for such interactions between quarks $q_1$, $q_2$ and $q_3$ involves the colour operator $d_{abc} T^a_{q_1} T^b_{q_2} T^c_{q_3}$, where $d_{abc}$ are the fully symmetrical algebra structure constants of SU($N$). For colour singlet baryons, three quarks are always in the antisymmetrical representation and, according to~(\ref{C3q}), 
\begin{equation}
{\cal C}_3=\frac{(N+2)(N+1)}{2N^2}.
\end{equation}
which is of order 1 at large-$N$ as required. 

The dynamics of three-quark interactions can be appraised through a generalisation of the model proposed in \cite{Pepin:2001is,Papp:2002xh}. We first chose a kinetic energy of the form 
\begin{equation}
	K(p)=\mu_\alpha \, p^\alpha+m ,
\end{equation}
that can accommodate for both massless ($\alpha=1$, $\mu_1=1$) and massive ($\alpha=2$, $\mu_2=\frac{1}{2m}$) quarks. We can expect that the most dramatic effect of three-quark interaction will manifest itself in confinement. So we set $V_3(x)={\cal C}_3 g_3\, x$ in agreement with the choice $U(x)=\sigma\, x$. In this case, one finds from (\ref{model0}) that an upper bound for the mass is given by
\begin{equation}\label{Mb}
	\frac{M}{N}=m+(1+\alpha)\left[ \frac{\mu_\alpha^{1/\alpha}}{\alpha}\sigma_3 \frac{Q}{N}\right]^{\alpha/(1+\alpha)},
\end{equation}
with 
\begin{equation}\label{Mb2}
\sigma_3=\sigma + \begin{pmatrix} N \\ 3 \end{pmatrix} \frac{{{\cal C}_3}}{N}g_3 \psi_3.
\end{equation}
Note that upper bounds for three massless quarks interacting via linear and Coulomb potentials are computed in the framework of the ET with relative errors around 15\% \cite{Semay:2015}. It appears clearly that the three-quark interaction can be absorbed in a shift of $\sigma$, this shift being of order 1 at large $N$: $\sigma_3\sim \sigma +\frac{\bar g_3}{2\sqrt 6}$. The mass formula (\ref{Mb}), although being approximate, predicts baryon Regge trajectories for light quarks, \textit{i.e.} $m=0$ and $\alpha=1$.

If $h > 3$, a great number of different structures are possible for the colour operator ${\cal C}_h$. All can be built from various combinations of the algebra structure constants $d_{abc}$ and $f_{abc}$. The equivalent of equation~(\ref{Mb}) is then obtained by replacing $\sigma_3$ by $\sigma_h$ with
\begin{equation}\label{Mb3}
\sigma_h=\sigma + \begin{pmatrix} N \\ h \end{pmatrix} \frac{{{\cal C}_h}}{N}g_h \psi_h.
\end{equation}
If we assume that for all values of $h$, ${\cal C}_h \sim 1$, as it is the case for $h=2$ and $3$, the string tension $\sigma$ is simply modified by a quantity $\sim a_h\,\sqrt{h(h-1)}\,\bar g_h$. Under this condition, the effect of multiquark interactions can always be absorbed in a redefinition of $\sigma$. The criterion ${\cal C}_h \sim 1$, being a sufficient condition to recover the expected baryon mass scaling at large $N$, may actually be seen as a selection criterion in model building: $h$-quark leading to a different scaling are ruled out.

\section{Baryon melting at finite temperature}\label{sec:deconf}

Up to now we focused on baryons at $T=0$. The existence of baryons above the deconfinement temperature $T_c$ has been suggested in \cite{Liao:2005hj}. Starting from a Hamiltonian with nonrelativistic kinetic energy and a finite-range two-body potential fitted on SU(3) lattice QCD, it has been shown that baryons may exist up to 1.6~$T_c$. May three (or more) quark interactions favour the binding of baryons above $T_c$? Although this topic deserves calculations that include properly the continuum sector, our framework may provide some indications. 

In a quark-gluon plasma, many colour channels are possible for a baryon. But ${\cal C}_2 \sim 1$ et ${\cal C}_3 \sim 1$ for all possible representations (see~\ref{sec:app}). Let us consider a nonrelativistic Hamiltonian with a particular $h$-quark interaction
\begin{equation}\label{Hamdef2}
	H=\sum^N_{i=1} \frac{\vec p_i^2}{2 m} - \vert {\cal C}_h\vert N^{1-h}\bar g_h\sum^N_{\{i_1,\dots,i_h\}} v_h(r_{\{i_1,\dots,i_h\}}),
\end{equation}
where $\bar g_h>0$ by definition of the arbitrary constant and where $v_h$ is a monotonic globally positive function with $v_h(x\to\infty)=0$ to mimic finite-range interactions above $T_c$. It is not relevant to consider massless quarks with short-range attractive potentials which cannot bind such particles. Moreover, it is physically expected that light quarks are dressed with a constituent mass by the surrounding bath of free quarks, antiquarks and gluons \cite{Liao:2005hj}.

Starting from (\ref{Hamdef2}), it can be obtained from \cite{Semay:2018rvp} that upper bounds of critical coupling constants -- the values of $\bar g_h$ below which a bound state with quantum number $Q$ melts -- for such potentials are given by
\begin{equation} \label{gmelt}
\bar g_h=\frac{Q^2}{2 m \vert{\cal C}_h\vert x_0^2 v_h(x_0)}\frac{N^h}{\begin{pmatrix} N \\ h \end{pmatrix}}\frac{h(h-1)}{N(N-1)} ,
\end{equation}
where the quantity $x_0^2 v_h(x_0)$ depends only of the form of $v_h$ and is determined by the relation $x_0 v'_h(x_0)+2 v_h(x_0)=0$. Note that the accuracy of the ET for nonrelativistic systems with two-body attractive Gaussian potentials has been tested as very good until $N = 20$ \cite{Semay:2015}. Since $Q^2\sim N^2$, it is readily seen that the criterion $\vert{\cal C}_h\vert\sim 1$ leads to $\bar g_h\sim 1$, hence to the conclusion that the melting temperature does not depend on $N$ at dominant order.

The more $\bar g_h$ is small, the more the interaction will lead to high-temperature for baryon melting since $\bar g_h$ is a priori a decreasing function of $T$ from renormalization arguments (see \textit{e.g.} \cite{Caswell} where the temperature $T$ is taken as the energy scale). Let us compare the critical constants for a given state ($Q$ and $m$ fixed) but for different $h$ and a common function $v_h$ $\forall\ h
$. Equation (\ref{gmelt}) leads to 
\begin{equation}
	\frac{\bar g_{h+1}}{\bar g_h}=\frac{\vert{\cal C}_{h}\vert}{\vert{\cal C}_{h+1}\vert}\frac{N}{N-h}\frac{(h+1)^2}{h-1}.
\end{equation} 
As an illustration one can write that, for baryons in a color singlet,
\begin{equation}
	\frac{\bar g_{3}}{\bar g_2}=9\frac{N^2}{N^2 -4}.
\end{equation} 
In other words, since $\bar g_3$ is well larger than $\bar g_2$, including three-quark interactions is not expected to significantly increase the baryons melting temperature.

\section{Summary}\label{sec:conclu}

The colour dependence of some possible many-quark interactions in baryons has been investigated in the framework of the large-$N$ approach of QCD with the quarks in the fundamental representation. Analytical information may be obtained below and above the deconfinement temperature $T_c$ with simple constituent quark models solved with the envelope theory.

Below $T_c$, baryons are bound states of $N$ quarks whose mass is proportional to $N$ as $N\to \infty$ ($M\sim N$). This sets constraints on the behaviour with $N$ of the possible colour $h$-body operators ${\cal C}_h$ that can be used in a constituent framework. Computations for $h=2$ and $3$ show that these operators behave as constants when $N\to \infty$ (${\cal C}_2 \sim 1$ and ${\cal C}_3 \sim 1$), which is compatible with $M\sim N$. Moreover, it is shown that $M\sim N$ remains true if ${\cal C}_h \sim 1$ for $h>3$. These are indications that the relevant structures for ${\cal C}_h$ with $h>3$ must be chosen in order that ${\cal C}_h \sim 1$. Our result suggest that $h-$quark interactions mostly result in a rescaling of one- and two-quark interaction parameters. 

Above $T_c$, our calculations show that the melting temperature of colourless baryons (if any) is independent of $N$ at leading order, and that the inclusion of multiquark interactions does not stabilize deconfined baryons.

\section*{Acknowledgments}
This work was supported by the Fonds de la Recherche Scientifique - FNRS under Grant Number 4.45.10.08.

\appendix

\section{Casimir operator for two and three quarks}\label{sec:app}

The eigenvalues $C^k_p$ of the SU($N$) Casimir operator of order $p$ for the totally antisymmetric and totally symmetric representations with $k$ indices are given respectively by \cite{Kobayashi:1973ey}
\begin{eqnarray}\label{cas}
C^k_p&=&\frac{k(N\pm 1)(N\mp k)}{2N^p(N\mp k\pm 1)}\left[(N\pm 1)^{p-1}(N\mp k)^{p-1}-(-k)^{p-1}\right],
\end{eqnarray}
with a normalisation such that $C^1_2=4/3$ for $N=3$, which is the usual normalisation in constituent models \cite{isgur,Buisseret:2010na,Buisseret:2011aa}.  Two quarks can only be in two colour representations: antisymmetric (A) which is the singlet one and symmetric (S). The corresponding values for ${\cal C}_{2\, \textrm{OGE}}$ are given by
\begin{eqnarray}\label{C2OGE}
{\cal C}_{2\, \textrm{OGE}}^{A}&=&-\frac{N+1}{2N}, \nonumber \\
{\cal C}_{2\, \textrm{OGE}}^{S}&=&\frac{N-1}{2N}. 
\end{eqnarray}
Both colour coefficients are of order 1 when $N\to \infty$. 

The procedure presented in \cite{Pepin:2001is} for the computation of the Casimir operator for three quarks in SU(3) can be generalised to SU($N$), and one has
\begin{equation}
{\cal C}_{3q}=d_{abc} T^a_{q_1} T^b_{q_2} T^c_{q_3}=\frac{1}{6}\left[C_3^{3q}-\frac{3}{2}\frac{N^2-4}{N}C^{3q}_2+6 C_3^{1q}\right],
\end{equation}
where $C_3^{3q}=d_{abc}\sum^3_{q_1,q_2,q_3=1}T^a_{q_1} T^b_{q_2} T^c_{q_3}=(C_3^3-C_2^3)/2$, $C_3^{1q}=d_{abc}T^a_{1} T^b_{1} T^c_{1}=(2C_3^1-C_2^1)/4$ and $C_2^{3q}=\delta_{ab}\sum^3_{q_1,q_2=1} T^a_{q_1} T^b_{q_2}=C_2^3/2$, and where the values $C_p^k$ are given in (\ref{cas}). Three quarks can only be in three colour representations: antisymmetric (A) which is the singlet one, symmetric (S) and mixed-symmetric (M). The corresponding values are given by
\begin{eqnarray}\label{C3q}
{\cal C}_{3q}^{A}&=&\frac{(N+2)(N+1)}{2N^2}, \nonumber  \\
{\cal C}_{3q}^{S}&=&\frac{(N-2)(N-1)}{2N^2}, \\
{\cal C}_{3q}^{M}&=&-\frac{N^2-4}{4N^2}.\nonumber 
\end{eqnarray}
For $N=3$, these values are in agreement with the ones obtained in \cite{Dmitrasinovic:2001nu,Pepin:2001is,Papp:2002xh}. All these colour coefficients are of order 1 when $N\to \infty$.

\end{document}